\documentclass[pre,aps,twocolumn,reprint,longbibliography]{revtex4-2}

\usepackage{graphicx}
\usepackage{amssymb,latexsym,amsthm,amsmath}    
\usepackage{mathtools}

\usepackage[draft]{changes}

\usepackage{color}
\usepackage{hhline}
\usepackage{colortbl}
\usepackage{ulem} 
\usepackage{xcolor}
\usepackage{hyperref}
\usepackage{soul}

\newcommand{\mb}[1]{\mbox{\bfseries \itshape #1}}

\begin{document}

\title{Flatness-based control of nonlinear complex networks}

\author{Irene Sendi\~na-Nadal, I. Leyva}
\affiliation{Complex Systems Group \& GISC, Universidad Rey Juan Carlos, 28933 M\'ostoles, Spain}

\author{Jean Pierre Barbot}
\affiliation{QUARTZ EA7393 Laboratory, ENSEA, 6 Avenue du Ponceau, 95014 Cergy-Pontoise, France}
\affiliation{LS2N, UMR 6005 CNRS, École Centrale de Nantes, 1 rue de la Noë, 44300 Nantes, France}

\author{Christophe Letellier}
\affiliation{Rouen Normandie University — CORIA, Avenue de l'Universit\'e, 
76800 Saint-Etienne du Rouvray, France}

\begin{abstract}
Effective control of nonlinear complex networks needs to identify not only the intervention sites but also the state variables to be monitored and the locations at which control inputs should be applied. Here, we present a network-scale generalization of an observability-guided, flatness-based control framework. The resulting control laws are implemented using sliding-mode control  and evaluated in heterogeneous networks of chaotic oscillators driven toward a prescribed periodic regime. We compare two control architectures: in the node-wise method, oscillators are independently sensed and actuated, while in the pairwise method, measurements acquired from a single node within each dyad are used to synthesize control inputs applied to both nodes. Across the parameter regimes considered, node-wise control generally achieves 
accurate control over a broader operating region, whereas pairwise control remains effective over a more restricted region while using half as many measurement locations for a given number of actuated nodes. Moreover, network interactions can enhance control by propagating the control influence to non-actuated nodes, creating a cooperative effect between the collective dynamics and the control action. Collectively, these findings quantify the trade-offs among measurement-location requirements, control performance, and control effort in nonlinear complex networks.

\end{abstract}

\maketitle

\begin{quotation} 
{\bf 
A network is not merely a collection of components to be controlled: its interactions can also help transmit the effect of the control action. We develop a strategy for controlling networks of nonlinear dynamical units while reducing the number of locations at which measurements must be collected. By measuring two state variables at one unit and using them to construct the control signals applied to a coupled pair, accurate control can be achieved without monitoring every actuated unit independently. The interactions also spread the control influence to units without direct actuation, creating a cooperative effect between the collective dynamics and the external intervention. This interplay provides a means of balancing measurement requirements, control accuracy, and energetic cost in nonlinear complex networks.
	}
\end{quotation}

\section{Introduction}

The control of collective dynamics in networked systems constitutes a major 
challenge across physics, engineering, and biology, with applications as relevant as 
power-grid stabilization and the suppression of cascading failures, the coordination of robotic swarms, or the reprogramming of cellular networks 
\cite{Liu2016,DSouza2023,Coraggio2026}. In many such systems, individual nodes 
display nonlinear dynamics; consequently, the control objective often extends 
beyond convergence to a singular point to the enforcement of a prescribed, 
time-dependent trajectory or dynamical regime. Representative examples include 
the regulation of biological and metabolic units \cite{Kim2009,Tang2012}, the synchronization of complex networks \cite{Yu13,Li15,Wu19}, and secure communications systems 
\cite{Pecora1990,Cuomo1993,Kethiri2026}. A key practical question is whether these collective behaviors can be realized while minimizing the number of nodes that require direct sensing and actuation and simultaneously reducing the associated energetic expenditure.

Controlling chaotic dynamics has been addressed through several approaches, including feedforward (non-feedback) control based on periodic excitation \cite{Kuz85}, the Ott–Grebogi–Yorke (OGY) method relying on linearization of the Poincaré map \cite{Ott1990}, the Pyragas time-delayed feedback technique \cite{Pyragas1992}, and nonlinear and adaptive control-engineering methods \cite{Jac95,Nij97}; for reviews, see \cite{Agu00,Fra06}. Extending these ideas to large networks of nonlinear units, however, requires addressing the additional challenges of sensor–actuator placement and network-mediated interactions. Seminal studies of network controllability \cite{Lin1974, Liu2011} proposed a structural framework to identify driver nodes, that is, nodes 
receiving time-varying external inputs capable of steering the dynamics of a complex directed network. These contributions clarified how the minimum number of driver nodes depends on the network’s degree distribution and motivated subsequent investigations into target control \cite{Gao2014}, the influence of nodes with low in-degree and out-degree \cite{Menichetti2014}, and the energetic constraints associated with controlling networks via a limited set of input channels \cite{Sun2013,Yan2012,Yan2015}. Nevertheless, such structural formulations typically model the network as a linear time-invariant system. Consequently, they generally do not guarantee the existence of an energy-efficient control solution, nor do they provide a control law that can enforce a predefined trajectory when the nodes exhibit inherently nonlinear dynamics.

Alternative methodologies that explicitly incorporate nonlinear network 
dynamics include compensatory perturbation frameworks \cite{Cornelius2013}, 
symmetry-based assessments of observability and controllability 
\cite{Whalen2015}, and numerical approaches, such as the one developed for networked systems with directed higher-order interactions \cite{Rizzello2026}.

A closely related research direction is pinning control, wherein an externally 
imposed reference trajectory is applied to a selected subset of nodes, and the remaining nodes are expected to converge via coupling-mediated 
interactions\cite{Li2004,Chen2007,Sorrentino2007,Porfiri2008,Almendral2009,Wang2002,Lin2016,Wu2012,Yuan2018,Tang2014}. Sufficient conditions for this 
convergence have been derived by
combining properties of the nodal dynamics with the spectrum of the Laplacian
restricted to the non-pinned nodes \cite{Porfiri2008}, or through the master stability function of the pinned network \cite{Sorrentino2007}. The performance of pinning control depends critically on both the cardinality and the topological placement of the pinned nodes. In particular, spectral analysis indicates that prioritizing high-degree nodes is beneficial when only a small fraction of nodes is directly controlled, whereas prioritizing low-degree nodes may be advantageous when the controlled fraction is larger \cite{Liu2021}.

Control techniques typically assume that the measured variables of each controlled node provide enough information to determine the control input. 
In networks of nonlinear oscillators, the selection of the measured output and the specific state equation into which the control input is applied are critical design choices, particularly under chaotic local dynamics. Indeed, the optimal 
strategy for reconstructing the states from partial measurements strongly 
depends on the intrinsic node dynamics and the chosen measured variables and 
thus cannot, in general, be deduced from network topology 
alone.\cite{Cowan2012,Gates2016,Zanhudo2017,Haber2018,Sendina2022}

This limitation can be alleviated by combining structural observability and Kalman’s duality principle \cite{Kal60} with the notion of differential flatness \cite{Letellier2021}. Specifically, graphical and symbolic analyzes can be employed to identify which measured variables allow the reconstruction of the state of a nonlinear oscillator \cite{Letellier2018,Let18,Sendina2022} and which state derivative provides an appropriate point for injecting the control input \cite{Let23b,Letellier2025ND}. Moreover, an observability path derived from a fluence graph makes it possible to represent both the system state and the control signal as functions of a suitably selected flat output and a finite set of its time derivatives, thereby making explicit the link between sensor–actuator placement and the governing equations.

Using this methodology, we previously derived flatness-based control laws for single chaotic oscillators and for dyads of diffusively coupled oscillators \cite{Letellier2023,Letellier2025Chaos,Letellier2025ND}. These controllers enable the enforcement of prescribed target dynamics and the induction of multiple forms of generalized synchronization, including configurations in which the coupled systems differ in state-space dimension and/or dissipation. Related flat-input formulations have also been investigated in the context of secure communications \cite{Nicolau2018}. Nevertheless, previous work was limited to isolated oscillators or a single coupled pair. Consequently, it remains unresolved whether flatness-based control can be extended to large, heterogeneous networks; how network topology and coupling structure influence control performance and energetic cost; and whether the observability properties of coupled subsystems can be leveraged to reduce the number of required measurement sites.

This study addresses these open questions by extending the observability-guided, flatness-based control framework from isolated oscillators and dyadic systems to complex networks of arbitrary sizes. We develop two complementary control architectures: a node-wise scheme, in which the locally optimal sensor–actuator configuration of a single oscillator is replicated independently across a selected subset of nodes, and a pairwise scheme, which extends the flat control law derived for a diffusively coupled dyad  \cite{Letellier2025ND} to a network-wide pairing strategy that exploits the joint observability of coupled oscillators, so that measurements from a single node of each pair suffice to synthesize the control inputs applied to both.

Applying both architectures to heterogeneous networks of diffusively coupled chaotic oscillators, we systematically assess how coupling strength, network heterogeneity, the controlled-node fraction, and node-selection criteria affect synchronization and control effort, quantifying the trade-off between reducing the number of measurement locations and preserving accurate, bounded control. In addition, we examine how control effects propagate from actuated to non-actuated nodes and how this propagation depends on degree heterogeneity and controller placement. The results indicate that network coupling can support indirect node control; however, this advantage is non-monotonic and dependent on the chosen control architecture. Moreover, the optimal degree-based node-selection strategy depends on the controlled-node fraction. Accordingly, the network functions not only as the target of control but also as a medium that transmits dynamical information and control influence.

The remainder of the paper is organized as follows. Section~II introduces the network model, the observability-guided sensor--actuator placement, the control strategies, and the performance measures. Section~III presents the results for heterogeneous networks and analyzes the network-mediated propagation of control. Section~IV summarizes our conclusions.

\section{Theoretical framework and performance measures}

\subsection{Network observability and sensor--actuator placement}\label{sec:observability}

This work builds upon previous developments in observability-based state reconstruction of nonlinear dynamical networks \cite{Sendina2022} and in flatness-based control of single and dyads of coupled chaotic oscillators \cite{Letellier2023,Letellier2025Chaos,Letellier2025ND}. Here, we investigate how observability-guided sensor placement and the distribution of control inputs affect the ability to steer the dynamics of large heterogeneous networks of coupled dynamical systems toward prescribed dynamics produced by an external driver oscillator. 

We consider a response network of $N$ coupled dynamical systems, where each 
node $i=1,\dots,N$ is described by an $m$-dimensional state vector 
$\mb{x}_i(t)$. The network dynamics is written as
\begin{equation}
\dot{\mb{x}}_i= \mb{f}(\mb{x}_i) -d \, \sum_{j=1}^{N} L_{ij}
\mb{q}(\mb{x}_j)
+
\mathbf{g}_i u_i,
\label{eq:network_general}
\end{equation}
where $\mb{f}$ defines the dynamics of each node, $\mb{q}$ is the coupling 
function, $\left(L_{ij}\right)$ is the Laplacian matrix, $d$ is the coupling 
strength, $u_i$ is a scalar control input, and $\mb{g}_i$ is a vector that 
specifies the state equation of node $i$ into which the input is injected. 
Although the framework applies to general nonlinear dynamical systems, we 
illustrate it using networks of diffusively coupled R\"ossler oscillators, with 
$\mb{f}(\mb{x}_i)=(-y_i-z_i,x_i+ a y_i, 2+z_i(x_i-4))$ and 
$\mb{q}(x,y,z)=(0,y,0)^T $ corresponding to diffusive coupling through the 
variable $y$. Unless otherwise stated, the R\"ossler parameter $a$ is set to 
$a = a_{\rm R} = 0.432$, for which the isolated response nodes evolve in a 
chaotic regime.

\begin{figure*}
    \centering
  \includegraphics[width=0.9\linewidth]{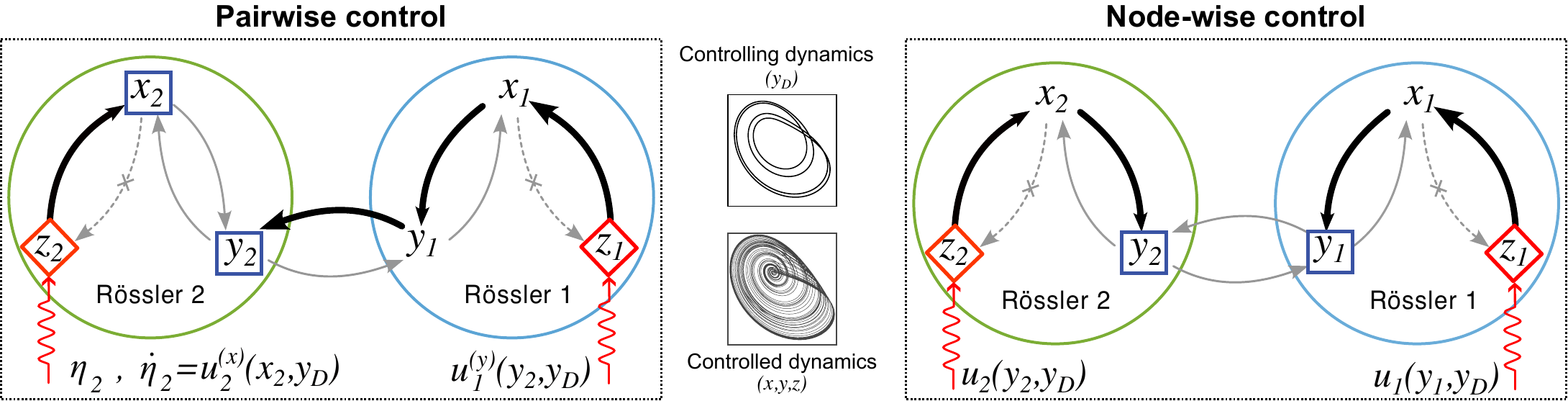}\\[5pt]
    \includegraphics[width=0.45\linewidth]{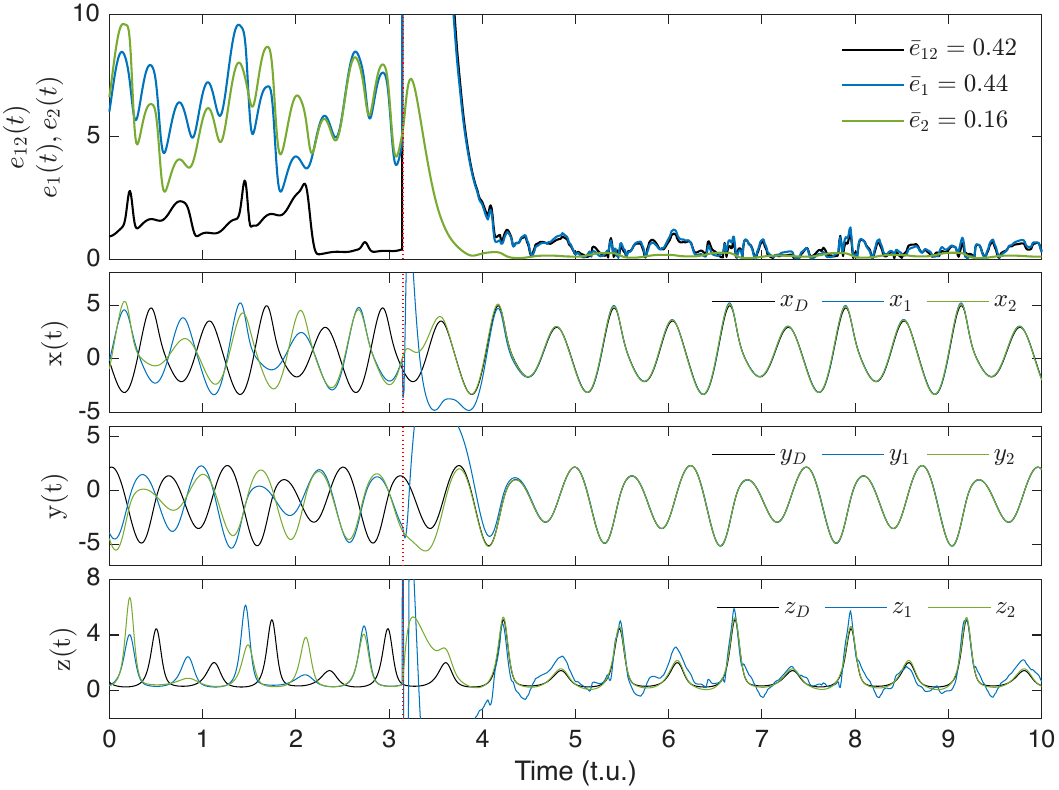}\hspace*{0.5cm}
   \includegraphics[width=0.45\linewidth]{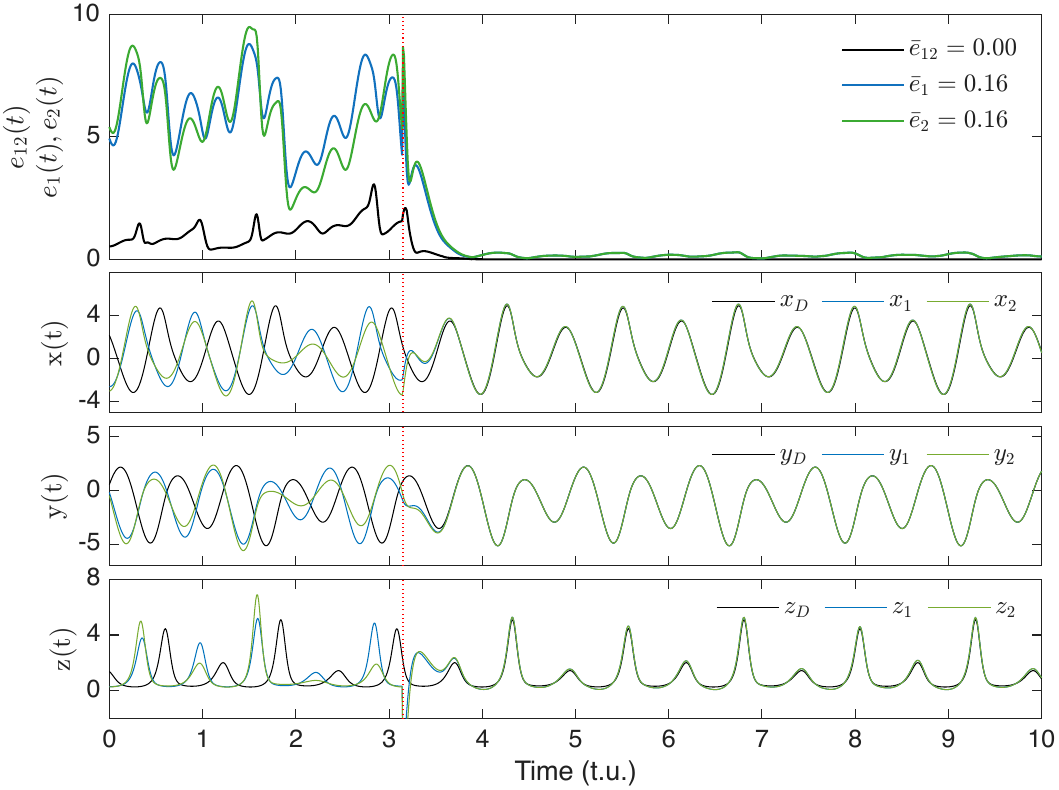}
     \caption{Comparison of pairwise (left) and node-wise (right) control for 
two diffusively coupled R\"ossler oscillators. The top panels show the 
corresponding control schemes and observability paths. In pairwise control, 
only variables $(x_2,y_2)$ of node 2, indicated by blue squares, are measured 
and used to construct the two control inputs. The input constructed from $x_2$ 
is implemented through a dynamic extension: the auxiliary variable $\eta_2$ is 
injected into the $\dot z_2$ equation and satisfies $\dot\eta_2 = u_2^{(x)}$. 
By contrast, in node-wise control, each controlled node is independently 
measured through its $y$ variable. The central panels show the periodic target
dynamics produced by the driver and the chaotic dynamics of the
uncontrolled response oscillators. The four lower panels show the
instantaneous synchronization and control errors, $e_{12}(t)$ and
$e_i(t)$, followed by the time series of the state variables $x$, $y$,
and $z$ for the driver and the two response oscillators. The vertical
dotted line marks the activation of the control, and the values
$\bar e_{12}$ and $\bar e_i$ reported in the first panel are time
averages evaluated after the control transient. The coupling strength
is $d=0.06$.
    \label{fig:pair-vs-node-schemes}}
\end{figure*}

The placement of sensors and control inputs is determined from the structural analysis of the node dynamics \cite{Liu2011,Liu2013,Liu2016,Sendina2022}. Following the graphical and symbolic observability approaches introduced in Ref.~\cite{Sendina2022}, each circle in the top right panel of Fig.\ref{fig:pair-vs-node-schemes} represents an individual R\"ossler node and contains its corresponding fluence graph. This graph is constructed directly from the node equations by distinguishing linear dependencies between state variables, represented by solid arrows, from nonlinear dependencies, represented by dashed arrows. Since nonlinear dependencies generally introduce singularities in the observability analysis, the nonlinear edge from $x$ to $z$, associated with the product $xz$, is discarded, as indicated by the crossed dashed arrows.
The resulting pruned graph contains the observability path \cite{Letellier2021}
\begin{equation}
z_i \rightarrow x_i \rightarrow y_i,
\label{eq:single-node-observability-path}
\end{equation}
highlighted by thick solid arrows. This path links the actuator and sensor locations: its endpoint identifies $y_i$ as the optimal observable variable (blue square), while its starting point identifies $\dot z_i$ as the corresponding controllable derivative (red diamond). Therefore, an individual R\"ossler oscillator can be controlled by measuring its $y$ variable and injecting the control signal into the corresponding $\dot z$ equation.

The same sensor-actuator configuration can be naturally replicated throughout a network by assigning the observability path in Eq. 
\eqref{eq:single-node-observability-path} to every controlled oscillator. Each 
controlled node is then independently observed through its own variable $y_i$ 
and controlled through its corresponding $\dot z_i$ equation, preserving the 
optimal sensor-actuator placement established for the isolated R\"ossler 
oscillator. Hereafter, we refer to this arrangement as 
\emph{node-wise control}.

The extension from one oscillator to two coupled oscillators, however, is not merely a duplication of the individual configuration. When two R\"ossler oscillators are diffusively coupled through their $y$ variables, the coupling itself becomes part of the fluence graph of the complete system. Previous observability studies showed that coupling through $y$ increases the observability of the dyad, making it possible to reconstruct the complete state of both oscillators from measurements performed on only one of them \cite{Sendina2022}. This result provides the key motivation for treating the two oscillators jointly rather than as two independent controlled systems.

The left panel of Fig.~\ref{fig:pair-vs-node-schemes} illustrates the resulting fluence graph of the coupled pair. Assuming that node 2 is selected as the measured node, the graph exhibits an observability path made of the two components, 
\begin{equation}
  z_2 \rightarrow x_2 \, , 
\label{eq:pair-direct-path}
\end{equation}
and
\begin{equation}
z_1 \rightarrow x_1 \rightarrow y_1 \rightarrow y_2.
\label{eq:pair-indirect-path}
\end{equation}
With these two components of the observability path, each vertex of the dyad’s fluence graph is visited exactly once \cite{Letellier2021}. They  are chosen because they allow measurements only at node 2. The first one corresponds to the directly observed oscillator through 
$x_2$ and $y_2$.
The second component extends through the coupling link $y_1\rightarrow y_2$, 
showing how information from the otherwise unobserved oscillator reaches the 
measured output. Consequently, measurements of the pair of variables 
$(x_2,y_2)$ provide sufficient information to reconstruct the dynamics of both 
oscillators. This enhanced observability constitutes the structural basis of 
the {\it pairwise control} strategy developed below.
For pairwise control, the network is decomposed into dyads following the 
hierarchical pairing procedure introduced in Ref.~\cite{Sendina2022}: nodes are 
paired starting from those with the lowest degree, and one node of each pair is designated as the measured node. Unpaired nodes, when present, remain 
uncontrolled.

\subsection{Flatness-based sliding-mode control}
\label{sec:flat_control}

The structural analysis introduced in Sec.~\ref{sec:observability} determines 
the variables to be measured and the state equations into which the control 
action is injected. For a flatness-based control law, these choices define the 
input--output configuration of the controlled system
\cite{Isidori1981,Isidori1995}. In the flatness-based approach, the
sensor and actuator are selected so that the measured outputs provide
global observability and the corresponding actuation channels provide
global controllability, while satisfying the additional differential
conditions required for flatness
\cite{Fliess1995,Letellier2021,Let23b}.

To formulate these conditions generically, we omit node indices and write the 
control-affine dynamics as $\dot{\mb{x}} = \mb{F}(\mb{x}) + \mb{g}(\mb{x}) u$,  
where $\mb{F}$ collects all terms independent of the control, including 
coupling terms when present, and $\mb{g}$ specifies the input vector field. Let 
$s = h(\mb{x})$ be the measured output. Denoting by 
$\mathcal{L}_{\mathbf v}\phi = \nabla \phi \cdot \mb{v}$ the Lie derivative of 
a scalar function $\phi(\mb{x})$ along a vector field $\mb{v}$, differentiation 
along a controlled trajectory gives $\dot\phi=\mathcal{L}_{\mathbf F}\phi +
\mathcal{L}_{\mathbf g}\phi \, u$.

The differential weight $w$ of the chosen input--output configuration 
(equivalent to the relative degree in classical nonlinear 
control\cite{Isidori1995}) is the smallest integer such that
\begin{equation}
    \mathcal{L}_{\mathbf g}
    \mathcal{L}_{\mathbf F}^{k}h=0,
    \qquad k=0,\ldots,w-2,
\label{eq:flat_condition_zero}
\end{equation}
while
\begin{equation}
    \mathcal{L}_{\mathbf g}
    \mathcal{L}_{\mathbf F}^{w-1} h \neq 0.
\label{eq:flat_condition_nonzero}
\end{equation}
Repeated differentiation then gives
$s^{(r)}=\mathcal{L}_{\mathbf F}^{r}h$ for $r=1,\ldots,w-1$, whereas
\begin{equation}
    s^{(w)}
    =
    \mathcal{L}_{\mathbf F}^{w}h(\mb{x})
    +
    \mathcal{L}_{\mathbf g}
    \mathcal{L}_{\mathbf F}^{w-1}h(\mb{x})\,u .
\label{eq:flat_input_output}
\end{equation}
Thus, $u$ appears explicitly for the first time in $s^{(w)}$. When the state 
and the input can also be reconstructed globally from $s$ and a finite number 
of its derivatives, $s$ is a flat output for the chosen input.

For an isolated R\"ossler oscillator, $\mb{F}$ reduces to the single-node 
vector field. Section~\ref{sec:observability} identified $y$ as the measured 
variable and the $\dot z$ equation as the location at which the control input 
is injected. Thus,
\begin{equation}
    h (\mb{x}) = y,
    \qquad
    \mb{g} = (0,0,1)^{\mathsf T},
\end{equation}
and
\begin{equation}
    \mathcal{L}_{\mathbf g}h=0,\qquad
    \mathcal{L}_{\mathbf g}\mathcal{L}_{\mathbf F} h = 0,\qquad
    \mathcal{L}_{\mathbf g}\mathcal{L}_{\mathbf F}^{2} h = -1.
  \label{eq:Rossler_flatness}
\end{equation}
The control input therefore does not appear in $\dot y$ or $\ddot y$, but 
enters explicitly in the third derivative,
\begin{equation}
    \dddot y = \mathcal{L}_{\mathbf F}^{3}h(\mb{x}) - u.
\label{eq:Rossler_input_output}
\end{equation}
The R\"ossler oscillator thus has differential weight $w=3$
for this input--output configuration. Moreover, its state can be
reconstructed globally from $y$ and its first two derivatives through
$x=\dot y-ay$ and $z=-y-\ddot y+a\dot y$. Hence, $y$ is a flat output
for an input applied to the $\dot z$ equation.

Different feedback laws can be built on this flat input--output configuration 
\cite{Fliess1995}. Here we use sliding-mode control, a nonlinear control 
strategy particularly suited to systems affected by model uncertainties and 
perturbations \cite{Perruquetti2002}. Sliding-mode techniques have also been 
successfully applied to the control and synchronization of chaotic systems
\cite{Singer1991,Konishi1998,Yau2000,Kocamaz2017}. 

To formulate this sliding-mode controller, we consider the tracking of the flat
output $s=h(\mathbf{x})$ introduced above and appearing in Eq.~\eqref{eq:flat_input_output}.
Let $s_D$ denote its target trajectory and define the corresponding
control error as $e_s = s - s_{\rm D}$. Since the control input $u$ first
appears explicitly in $s^{(w)}$, a sliding variable is constructed
from $e_s$ and its successive derivatives up to order $w-1$:
\begin{equation}
    \sigma =
    \sum_{k=0}^{w-1} \kappa_k e_s^{(k)} =
    \sum_{k=0}^{w-1} \kappa_k \, \left(s^{(k)}-s_{\rm D}^{(k)}\right).
\label{eq:sliding_manifold}
\end{equation}
On the sliding manifold $\sigma=0$, Eq.~\eqref{eq:sliding_manifold} defines a 
linear differential equation of order $w-1$ for the control error. The 
coefficients $\kappa_k$ are chosen so that all roots of its characteristic 
polynomial are located at $-\lambda$, with $\lambda>0$. 
Equivalently, $\kappa_k$ are the coefficients of the characteristic
polynomial
\begin{equation}
    P(p)=(p+\lambda)^{w-1}.
\end{equation}
Since this polynomial is monic, the coefficient of the highest-order
error derivative in Eq.~\eqref{eq:sliding_manifold} is
$\kappa_{w-1}=1$.

To drive the trajectory toward the sliding manifold, we impose the
homogeneous finite-time dynamics
\begin{equation}
    \dot{\sigma} = -K_g\,\operatorname{sgn}(\sigma)|\sigma|^{\chi},
\label{eq:reaching_law}
\end{equation}
where $K_g > 0$ determines the control strength and $\chi$ is the homogeneity 
degree of the sliding dynamics \cite{Bernuau2014}. Since $\dot{\sigma}$ 
contains $\kappa_{w-1}s^{(w)}$, substituting Eq.~\eqref{eq:flat_input_output} 
into Eq.~\eqref{eq:reaching_law} and solving for $u$ yields the corresponding 
sliding-mode control law. The coefficient multiplying $u$ in $\dot{\sigma}$ is
$\kappa_{w-1} \mathcal{L}_{\mathbf g}\mathcal{L}_{\mathbf F}^{w-1}h$,
which is nonzero by the flatness condition \eqref{eq:flat_condition_nonzero}.

The resulting exact control law contains, in addition to the homogeneous term 
in Eq.~\eqref{eq:reaching_law}, terms arising from the intrinsic dynamics and 
from the successive control errors. These terms remain bounded along bounded 
trajectories and can be regarded as perturbations of the homogeneous sliding 
dynamics.  For a sufficiently large $K_g$, their contribution becomes small
relative to the homogeneous term, and the actuating signal can be approximated 
as
\begin{equation}
    u \simeq -\frac{K_g} {\kappa_{w-1}
     \mathcal{L}_{\mathbf g} \mathcal{L}_{\mathbf F}^{w-1}h}
    \operatorname{sgn}(\sigma)|\sigma|^{\chi}.
\label{eq:simplified_sliding_control}
\end{equation}
This approximation preserves finite-time convergence to a neighborhood of the 
sliding manifold, whose size decreases as the ratio between the bounded 
perturbation and $K_g$ decreases \cite{Bernuau2014,Letellier2025ND}.

The target signal $s_D$ is produced by an external driver
\begin{equation}
	\dot{\mb{x}}_{\rm D} = \mb{f}_{\rm D} (\mb{x}_{\rm D}).
\label{eq:driver}
\end{equation}
In the simulations reported below, the driver is a R\"ossler oscillator 
governed by the same vector field $\mb{f}$ as the response nodes, but operating 
in a stable period-4 regime with $a = a_{\rm D} = 0.38$. The control objective 
is therefore not merely to synchronize the network, but rather to suppress the 
intrinsic chaotic dynamics of the response oscillators and drive them towards a 
prescribed periodic orbit. Equation~(\ref{eq:simplified_sliding_control}) 
provides the common basis to implement the specific sliding manifolds and 
actuating signals for the node-wise and pairwise control strategies described 
next. 

\subsection{Node-wise and pairwise control strategies}
\label{sec:node-and-pair-control-laws}

We consider a network of diffusively $y$-coupled R\"ossler oscillators driven 
by a R\"ossler system, implementing the general framework introduced above 
through two strategies: node-wise and pairwise control. 
Figure~\ref{fig:pair-vs-node-schemes} illustrates both strategies for a pair of 
coupled R\"ossler oscillators, which serves to introduce their specific sliding 
manifolds and actuating signals before extending them to larger networks.

In the node-wise strategy, each controlled oscillator is treated independently. 
As established in Sec.~\ref{sec:observability}, $y_i$ is measured, and the 
control input $u_i$ is applied to $\dot z_i$, yielding a differential weight 
$w = 3$. Using Eq.~\eqref{eq:sliding_manifold} with coefficients 
$\kappa_0 = \lambda^2$, $\kappa_1 = 2 \lambda$, and $\kappa_2 = 1$, the 
corresponding sliding manifold is
\begin{equation}
	\sigma_i = \lambda^2 (y_i - y_{\rm D}) 
	+ 2 \lambda (\dot y_i - \dot y_{\rm D}) + (\ddot y_i - \ddot y_{\rm D}),
\label{eq:sigma-node}
\end{equation}
and Eq.~\eqref{eq:simplified_sliding_control} gives
\begin{equation}
  u_i = K_g\, \operatorname{sgn}(\sigma_i)|\sigma_i|^{\chi},
\label{eq:node-control}
\end{equation}
where $\kappa_{w-1}=\kappa_2=1$ and
$\mathcal{L}_{\mathbf g}\mathcal{L}_{\mathbf F}^{w-1} h
=\mathcal{L}_{\mathbf g}\mathcal{L}_{\mathbf F}^{2} h = -1$. Each controlled 
node therefore receives an independent actuating signal determined from its 
local control error and its derivatives.

In the pairwise strategy, the two coupled oscillators are treated jointly. Let 
$(i,j)$ denote a controlled pair, with $j$ designating the measured node and 
$i$ its non-measured partner. As discussed in Sec.~\ref{sec:observability}, the 
state of the dyad can be reconstructed from measurements of $x_j$ and $y_j$. 
The corresponding observability paths,
\[
z_i\rightarrow x_i\rightarrow y_i \rightarrow y_j,
\qquad
z_j\rightarrow x_j \, , 
\]
not only do the differential lengths differ, but more importantly, the additional existence of the path $z_j\rightarrow x_j \rightarrow y_j$ means that computing successive Lie derivatives of $y_j$ leads to reaching variable $z_j$
before its counterpart $z_i$, preventing a compatible multivariable flatness 
construction, since reconstructing $z_i$ cannot be performed independently of
reconstructing $z_j$. 

To overcome this mismatch, the shorter path is dynamically extended: an 
auxiliary variable $\eta_j$ is introduced into the $\dot z_j$ equation, and the
corresponding control input is defined through $\dot\eta_j = u_j^{(x)}$ 
\cite{Letellier2025ND}. This extension delays the explicit appearance of 
$u_j^{(x)}$ by one derivative. The resulting flat outputs $h_y = y_j$ and 
$h_x = x_j$ then have differential weights $w_y = 4$ and $w_x = 3$, 
respectively.

For the first control loop, based on the flat output $y_j$ with $w_y = 4$,
the sliding manifold is
\begin{equation}
  \begin{array}{rl}
	  \sigma_j^{(y)} = & \lambda^3 (y_j - y_{\rm D})
	  + 3 \, \lambda^2 (\dot y_j - \dot y_{\rm D}) \\[0.1cm]
	  & + 3 \, \lambda (\ddot y_j - \ddot y_{\rm D})
	+ (\dddot y_j - \dddot y_{\rm D}).
  \end{array}
\label{eq:sigma-pair-y}
\end{equation}
The corresponding actuating signal $u_i^{(y)}$ is applied directly
to the $\dot z_i$ equation. Since its effect reaches the measured
output $y_j$ through the diffusive coupling between the two nodes,
\[
  \mathcal{L}_{\mathbf g_i}
  \mathcal{L}_{\mathbf F}^{3}h_y=-d.
\]
Equation~\eqref{eq:simplified_sliding_control} therefore reduces to
\begin{equation}
  u_i^{(y)} = \frac{K_g}{d}\,
  \operatorname{sgn} \! \left( \sigma_j^{(y)} \right)
  \left| \sigma_j^{(y)} \right|^\chi .
\label{eq:pair-control-y}
\end{equation}

This coupling-mediated law is defined only for $d > 0$ and becomes singular as 
$d \to 0$. At $d = 0$, the two nodes of each selected pair are instead 
controlled independently through their respective $y$-based node-wise laws, 
without the dynamic extension, thus providing an uncoupled reference. The 
consequences of the $1/d$ amplification in the weak-coupling regime are 
discussed below.

For the second control loop, based on the flat output $x_j$ with $w_x = 3$,
a target $x$-signal consistent with the response dynamics is reconstructed from 
the driver output as
\[ x_{\rm D}^{(R)} = \dot y_{\rm D} - a_{\rm R} y_{\rm D}, \]
where $a_{\rm R}$ is the response parameter. The corresponding sliding
manifold is
\begin{equation}
  \sigma_j^{(x)} =
	\lambda^2 \left( \displaystyle x_j - x_{\rm D}^{({\rm R})} \right)
	+ 2 \lambda \left( \displaystyle \dot x_j - \dot x_{\rm D}^{({\rm R})}
	\right)
	+  \left( \displaystyle \ddot x_j - \ddot x_{\rm D}^{({\rm R})}
	\right).
\label{eq:sigma-pair-x}
\end{equation}
For the dynamically extended input,
$\mathcal{L}_{\mathbf g_j}\mathcal{L}_{\mathbf F}^{2} h_x = -1$, and
Eq.~\eqref{eq:simplified_sliding_control} becomes
\begin{equation}
  u_j^{(x)} = K_g\,
  \operatorname{sgn} \! \left( \sigma_j^{(x)} \right)
  \left| \sigma_j^{(x)} \right|^\chi .
\label{eq:pair-control-x}
\end{equation}
Thus, both pairwise control signals are constructed from measurements at node 
$j$. The signal $u_i^{(y)}$ acts directly on the $\dot z_i$ equation of its 
non-measured partner $i$ through the coupling-mediated $y$ channel, whereas 
$u_j^{(x)}$ drives the auxiliary dynamics $\dot\eta_j$ and therefore acts on 
the measured node $j$ through the term $\eta_j$ entering its $\dot z_j$ 
equation.  For the particular dyad shown in the left panel of 
Fig.~\ref{fig:pair-vs-node-schemes}, $i=1$ is the non-measured node and $j = 2$ 
is the measured node, as indicated by the blue squares marking $x_2$ and $y_2$.

Note that the control laws above are implemented in a decentralized manner: 
each actuating signal is computed from the local error of its own node or pair, 
and neither the inputs applied to other neighbours nor, in the pairwise scheme, 
the crossed contributions are taken into account.\cite{Letellier2025ND} This 
choice keeps each controller local and independent of the global network 
structure.

\subsection{Performance metrics}

To quantify the performance of the control strategies, we consider separately 
the ability of the response network to follow the target dynamics and the 
degree of synchronization among its nodes. 

For each response node $i$, we define the instantaneous control error with 
respect to the driver as
\begin{equation}
  e_i(t) =
	\left\| \mb{x}_i (t) - \mb{x}_{\rm D} (t) \right\|_2,
\label{eq:tracking_error}
\end{equation}
where $\mb{x}_i (t) = [x_i (t), y_i (t), z_i (t)]^{\mathsf T}$, 
$\mb{x}_{\rm D} (t) = [x_{\rm D} (t),y_{\rm D} (t),z_{\rm D} (t)]^{\mathsf T}$
and $\|\cdot\|_2$ denotes the Euclidean norm. The instantaneous synchronization 
error between nodes $i$ and $j$ is similarly defined as 
\begin{equation}
  e_{ij}(t) =
  \left\| \mb{x}_i (t) - \mb{x}_j (t) \right\|_2 .
\label{eq:sync_error}
\end{equation}
 
Although the control laws are designed using measured outputs and 
their derivatives, the control and synchronization errors are evaluated in the 
full state space. This provides a demanding assessment of control performance, 
since accurate control of the measured variables does not necessarily imply an 
equally accurate reconstruction of all state variables.

Observe that, since ($a_{\rm D} \neq a_{\rm R}$), the control error $e_i(t)$ 
cannot vanish even under ideal control. In our setting, imposing  
$y_i(t) = y_{\rm D}(t)$ means that perfect tracking of the measured output 
yields 
\[ 
  \begin{array}{rl} 
    e_i^{\star}(t) & = \| \mb{x}_i(t) - \mb{x}_{\rm D}(t) \|_2 \\
	& = |a_{\rm R} - a_{\rm D} |\, \sqrt{y_{\rm D}^2(t) 
	  + \dot y_{\rm D}^2(t)} \\
	  & \simeq 0.16 \,  .
  \end{array}
\]
This residual error, which stems from forcing a mismatched 
dynamics through a single actuation channel, represents the best achievable 
performance for this sensor-actuator configuration, consistent with the 
node-wise control results in Fig.~\ref{fig:pair-vs-node-schemes}.

The time series shown in Fig.~\ref{fig:pair-vs-node-schemes} illustrate these 
two errors for the minimal configuration of two coupled R\"ossler oscillators, 
together with the corresponding state variables. The first panel for each 
control strategy shows $e_{12} (t)$ and the two control errors $e_1 (t)$ and 
$e_2 (t)$, while the remaining three panels show the time series of $x$, $y$, 
and $z$ for the driver and the two response oscillators. Before control 
activation, the response oscillators evolve chaotically and remain separated 
from the periodic target trajectory. Once control is activated, the control 
errors decrease and the state variables approach those of the driver. Under 
node-wise control, both oscillators are independently measured and controlled 
and consequently exhibit nearly identical residual control errors 
$\bar e_i \simeq \bar e^{\star}$. Under 
pairwise control, instead, both actuating signals are constructed from 
measurements performed on a single node of the pair. The measured node closely 
follows the driver, whereas the non-measured node retains a larger residual 
error, most clearly visible in the $z$ variable. The values $\bar e_i$ and 
$\bar e_{12}$ reported in the figure correspond to time averages evaluated 
after the control transient.

For a network of $N$ nodes, the instantaneous global control error is defined 
by averaging the node control errors,
\begin{equation}
  C(t) = \frac{1}{N} \sum_{i=1}^{N} e_i (t) \, ,
\label{eq:control_error}
\end{equation}
while the intra-network synchronization error is the mean distance over all 
distinct pairs of nodes,
\begin{equation}
  S(t) = \frac{2}{N(N-1)} \sum_{i<j} e_{ij} (t).
\label{eq:network_sync}
\end{equation}
Thus, $C (t)$ measures the distance of the response network from the target 
dynamics, whereas $S (t)$ measures the dispersion of the response trajectories 
independently of the target.

The control effort is quantified from the squared amplitudes of the control 
signals. Let $\mathcal{C}$ denote the set of controlled nodes in the node-wise 
strategy and $\mathcal{P}$ the set of controlled pairs $(i,j)$ in the pairwise 
strategy, where $i$ and $j$ denote the non-measured and measured nodes, 
respectively. The instantaneous mean control energy per network node is defined 
for each strategy as
\begin{equation}
  U (t)= 
  \left\{
    \begin{array}{l}
        \displaystyle
        \frac{1}{N} \sum_{i \in \mathcal{C}} u_i^2 (t) \\[-1mm]
\multicolumn{1}{r}{\text{for node-wise control},}
\\[3mm]
\displaystyle
\frac{1}{N}
\sum_{(i,j)\in\mathcal{P}}
\left[
\left(u_i^{(y)}(t)\right)^2+
\left(u_j^{(x)}(t)\right)^2
\right]
\\[-1mm]
\multicolumn{1}{r}{\text{for pairwise control}.}
\end{array}
\right.
\label{eq:control_energy}
\end{equation}

The three performance measures are averaged over a time window of
duration $T$, after discarding the transient regime, according to
\begin{equation}
M=
\frac{1}{T}
\int_{t_0}^{t_0+T} M(t)\,dt,
\qquad M\in\{C,S,U\}.
\label{eq:time_averaged_metrics}
\end{equation}

\begin{figure}
    \centering
    \includegraphics[width=\linewidth]{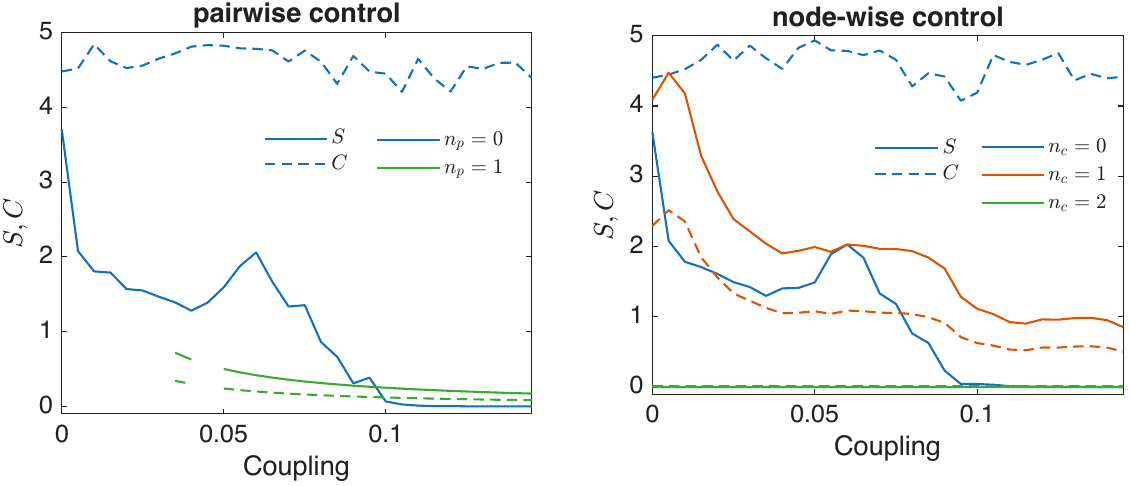}
    \caption{Control and synchronization errors for two diffusively coupled 
R\"ossler oscillators as a function of the coupling strength $d$, for pairwise 
(left) and node-wise (right) control.  Solid lines show the intra-network 
synchronization error $S$, whereas dashed lines show the control error $C$. For 
pairwise control, $n_{\rm p} = 0$ denotes the uncontrolled dyad and 
$n_{\rm p} = 1$ a single controlled pair. For node-wise control, $n_{\rm c} = 
0,1,2$ denotes the number of independently controlled nodes.}
    \label{fig:pair_synvscon}
\end{figure}

Before extending these quantities to large networks, 
Fig.~\ref{fig:pair_synvscon} illustrates their behavior for the minimal 
two-node configuration. The synchronization error $S$ (solid lines) and the 
control error $C$ (dashed lines) are shown as functions of the coupling 
strength $d$. In the absence of control ($n_{\rm c} = 0$ or $n_{\rm p} = 0$), 
the two strategies reduce to the same uncontrolled dyad. Increasing $d$ 
progressively promotes synchronization between the two response oscillators, as 
reflected by the decrease of $S$, while $C$ remains large and almost constant 
since no control is applied.

For node-wise control, the case $n_{\rm c} = 1$ corresponds to controlling only 
one oscillator, whereas for $n_{\rm c} = 2$ both oscillators are independently 
measured and controlled. The latter configuration produces the smallest 
control and synchronization errors. Pairwise control with $n_{\rm p} = 1$ also 
acts on the complete dyad, but uses measurements from only one of its nodes to 
construct the two actuating signals. Consequently, both $C$ and $S$ remain 
larger than for node-wise control with $n_{\rm c} = 2$. This difference 
quantifies the trade-off inherent to the pairwise strategy: reduced sensing 
requirements are obtained at the expense of lower control accuracy.

A distinctive limitation of pairwise control also appears at weak coupling. As 
shown in Sec.~II\,C, the actuating signal applied to the non-measured node 
contains the inverse coupling factor $1/d$, since $\mathcal{L}_{\mathbf{g}_i}\mathcal{L}_{\mathbf{F}}^{3} h_y = -d$.
Therefore, decreasing $d$ amplifies the required control signal and, for 
sufficiently weak coupling, may drive the response trajectory outside its 
bounded orbit. A finite coupling strength is thus required in the pairwise 
strategy both to transmit information between the two nodes and to maintain a 
bounded control action.

To facilitate comparison between the two strategies on a common scale, pairwise 
control is hereafter parameterized by the equivalent number of controlled 
nodes, $n_{\rm c} = 2 \, n_{\rm p}$, where $n_{\rm p}$ denotes the number of 
controlled pairs.

For comparisons across coupling strengths and numbers of controlled nodes, we 
finally introduce normalized performance measures. The normalized control error 
is defined as
\begin{equation}
	\widehat{C}(n_{\rm c},d) = \frac{C (n_{\rm c},d)}{C_0}, \qquad 
	C_0 = C(n_{\rm c},0),
\label{eq:norm_control}
\end{equation}
where $C_0$ is the control error obtained for the same control configuration 
when the response nodes are uncoupled. Similarly, the normalized control 
energy is
\begin{equation}
	\widehat{U} (n_{\rm c},d) = \frac{U (n_{\rm c},d)}{U_0}, \qquad
	U_0 = U(n_{\rm c},0).
\label{eq:norm_energy}
\end{equation}
Thus, $\widehat C$ and $\widehat U$ quantify how network coupling modifies, 
respectively, the control performance and the control energy relative to the 
same set of controlled nodes operating without network coupling.

The synchronization error requires a different reference because 
synchronization itself is induced by the network coupling. We therefore define
\begin{equation}
	\widehat{S}(n_{\rm c},d) = \frac{S(n_{\rm c},d)}{S_0}, \qquad
  S_0 = S(0,d),
\label{eq:norm_sync}
\end{equation}
where $S_0$ is the synchronization error of the uncontrolled network at the 
same coupling strength. Hence, $\widehat S$ isolates the additional effect of 
the control action on the synchronization already produced by the network 
interactions. Values $\widehat C < 1$ and $\widehat U < 1$ indicate that 
coupling improves control or reduces the required control energy, 
respectively, whereas $\widehat S < 1$ indicates that control enhances 
synchronization relative to the uncontrolled network.


\begin{figure*}[th]
  \includegraphics[width=0.8\textwidth]{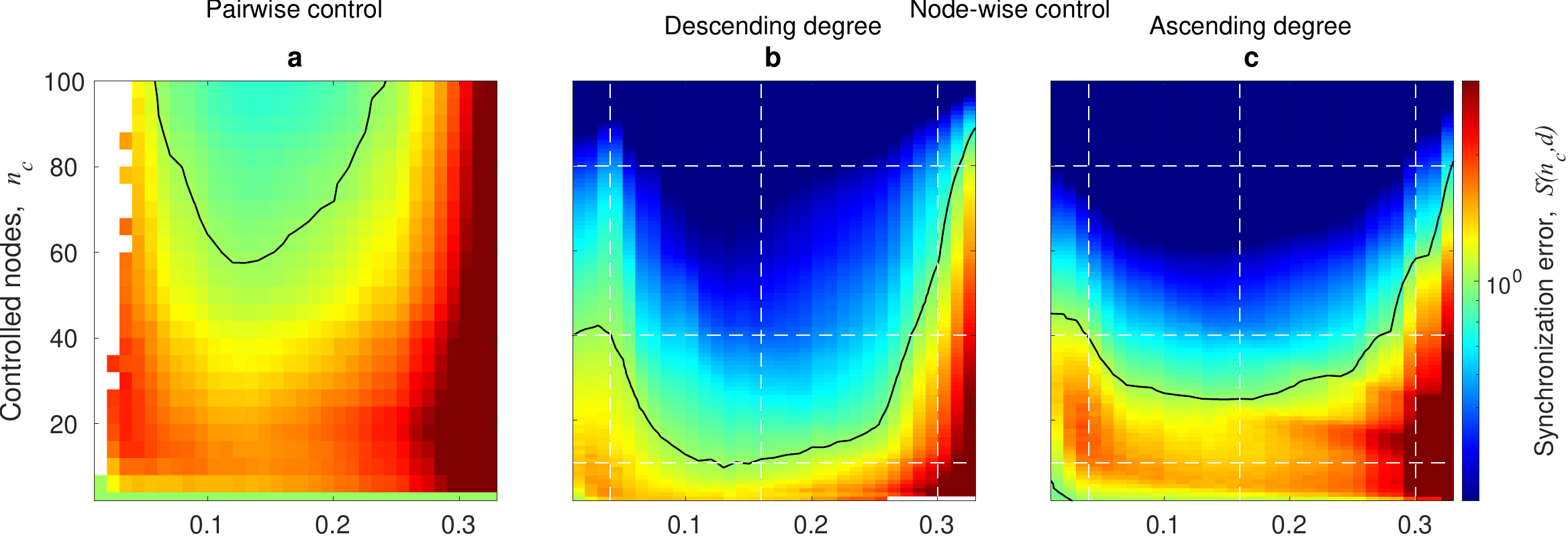}
 \includegraphics[width=0.8\textwidth]{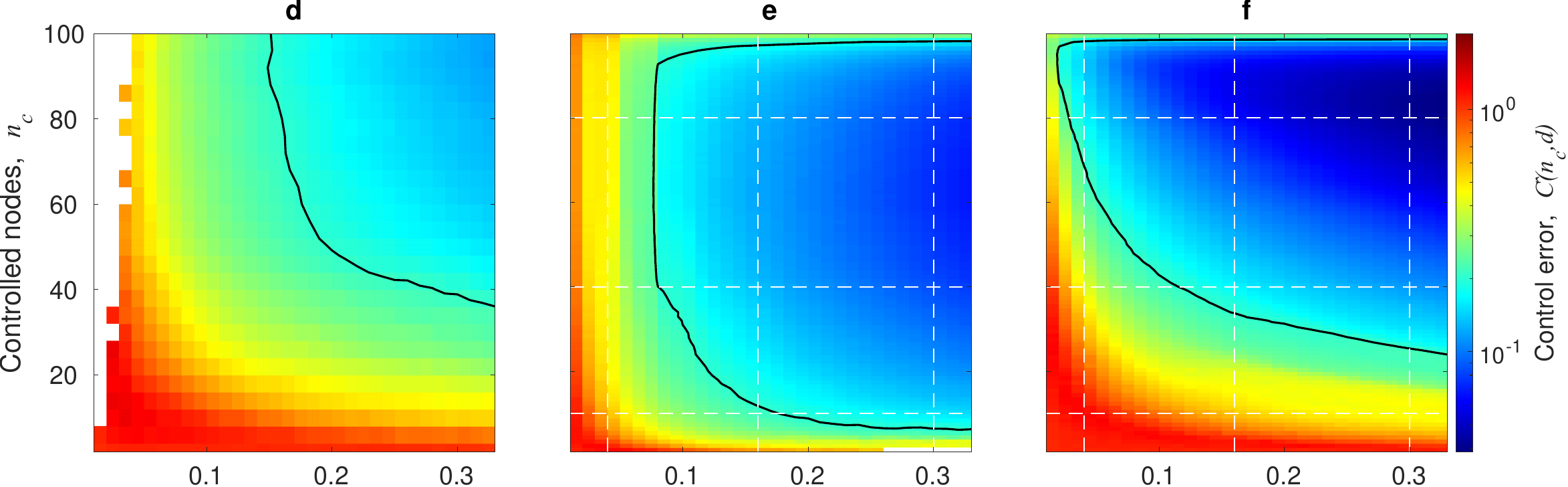}
 \includegraphics[width=0.8\textwidth]{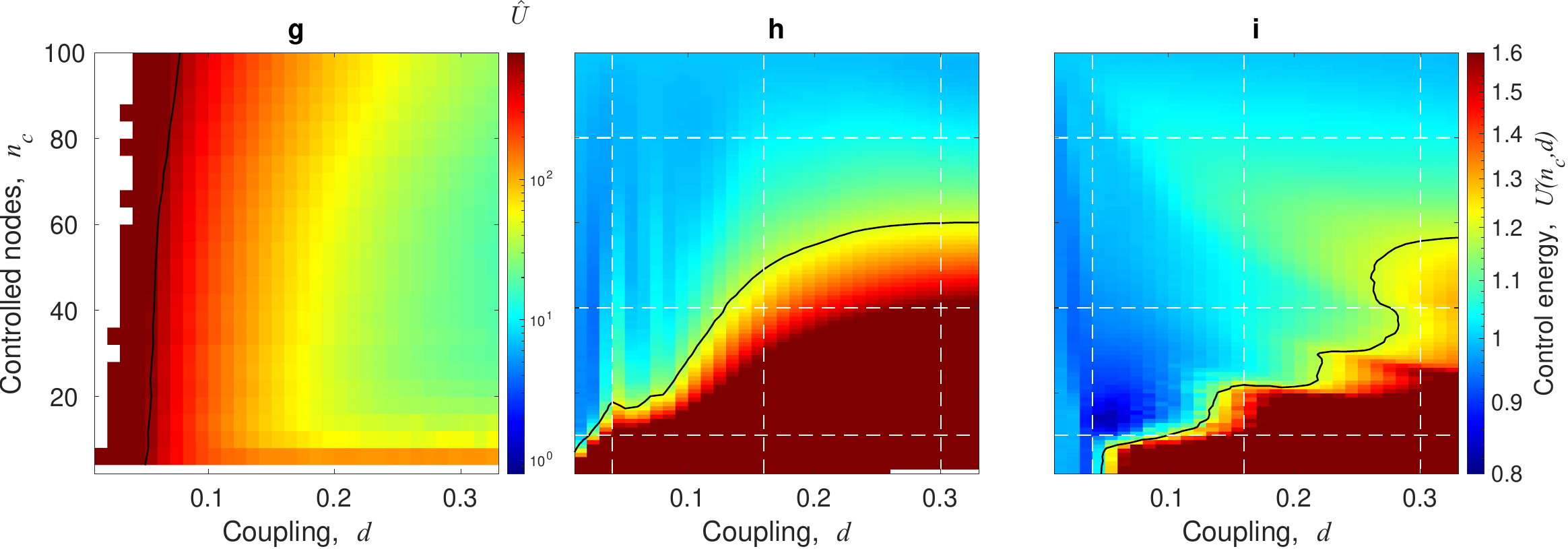}
  \caption{Comparison of pairwise and node-wise control in networks with 
heterogeneous degree distribution, with $N=100$ and $\langle k \rangle = 4$. 
The logarithmic color maps show the normalized synchronization error 
$\widehat{S} (n_{\rm c},d) = S (n_{\rm c},d)/S(0,d)$ (top row), the normalized 
control error $\widehat{C}(n_{\rm c},d) = C(n_{\rm c},d)/C(n_{\rm c},0)$ 
(middle row), and the normalized control energy $\widehat{U} (n_{\rm c},d)
= U (n_{\rm c},d)/U(n_{\rm c},0)$ (bottom row), as functions of the coupling 
strength $d$ and the number of controlled nodes $n_{\rm c}$. The left column 
[(a), (d), and (g)] corresponds to pairwise control, with pairs selected 
according to the hierarchical pairing order. The central [(b), (e), and (h)] 
and right [(c), (f), and (i)] columns correspond to node-wise control with 
nodes selected in descending and ascending degree order, respectively. For 
pairwise control, the vertical coordinate represents the equivalent number of 
controlled nodes, $n_{\rm c} = 2 \, n_{\rm p}$. Black contour lines indicate 
$\widehat{S}=1$ in the top row, $\widehat{C}=0.2$ in the middle row, and the 
strategy-dependent energy thresholds $\widehat{U} = 100$ for pairwise control 
and $\widehat{U} = 1.2$ for node-wise control in the bottom row. White regions 
correspond to parameter combinations for which the controlled trajectories 
become unbounded and the performance indicators cannot be evaluated. White 
dashed lines indicate the values of $n_{\rm c}$ and $d$ selected for the 
one-dimensional sections shown in Fig.~\ref{fig:cutscolormaps}. Results are 
averaged over $25$ independent network realizations and $5$ initial conditions 
per realization. Control parameters are $K_g = 20$, $\lambda = 1$, and 
$\chi = 0.7$.
  \label{fig:colormaps}
  }
\end{figure*}

\section{Results}
\label{sec:results}

Having introduced the control schemes and the associated performance metrics in the minimal two-node configuration, we now investigate their behavior in large heterogeneous networks.

\subsection{Control performance in heterogeneous networks}

We first investigate the performance of the pairwise and node-wise
control strategies in ensembles of heterogeneous networks produced
with the Barab\'asi--Albert (BA) growth mechanism \cite{Barabasi1999}. The
response networks have a size $N = 100$ and a mean degree
$\langle k \rangle = 4$. For each coupling strength $d$ and each
prescribed control level, a separate simulation is performed with a
fixed set of controlled nodes. These sets are tested as the control
level is increased. In the node-wise strategy, the set corresponding
to $n_{\rm c}$ is formed by selecting the first $n_{\rm c}$ nodes according to
descending degree, ascending degree, or a random ordering. In the
pairwise strategy, the set corresponding to $n_{\rm p}$ is formed by
selecting the first $n_{\rm p}$ dyads according to the hierarchical pairing
procedure introduced in Sec.~\ref{sec:observability}. Since each
controlled pair contains two actuated nodes, pairwise results are
represented in terms of the equivalent number of controlled nodes,
$n_{\rm c} = 2 n_{\rm p}$.

The equations are integrated using a fourth-order Runge-Kutta scheme with time 
step $\Delta t = 0.01$. The time derivatives 
computed for producing the actuating signal are 
evaluated numerically using centered finite-difference formulas over a 
four-sample window. All the time lengths are expressed in terms of the characteristic oscillation period $T_{\rm R}=2\pi$ of an isolated, uncontrolled 
R\"ossler oscillator. After initialization, the uncontrolled network 
dynamics is allowed to evolve for $20 \, T_{\rm R}$. For configurations with 
$n_{\rm c} > 0$, the control is activated, and an additional transient of 
$20 \, T_{\rm R}$ is discarded before the performance metrics are averaged over 
the following $100 \, T_{\rm R}$. For the uncontrolled baseline, 
$n_{\rm c} = 0$, the metrics are averaged over $100 \, T_{\rm R}$ immediately 
after the initial uncontrolled transient. The reported results are additionally 
averaged over $25$ independent network realizations and $5$ initial conditions 
per network. Unless otherwise stated, the control parameters are $K_g = 20$, 
$\lambda = 1$, and $\chi = 0.7$.

\begin{figure*}[t]
    \includegraphics[width=\linewidth]{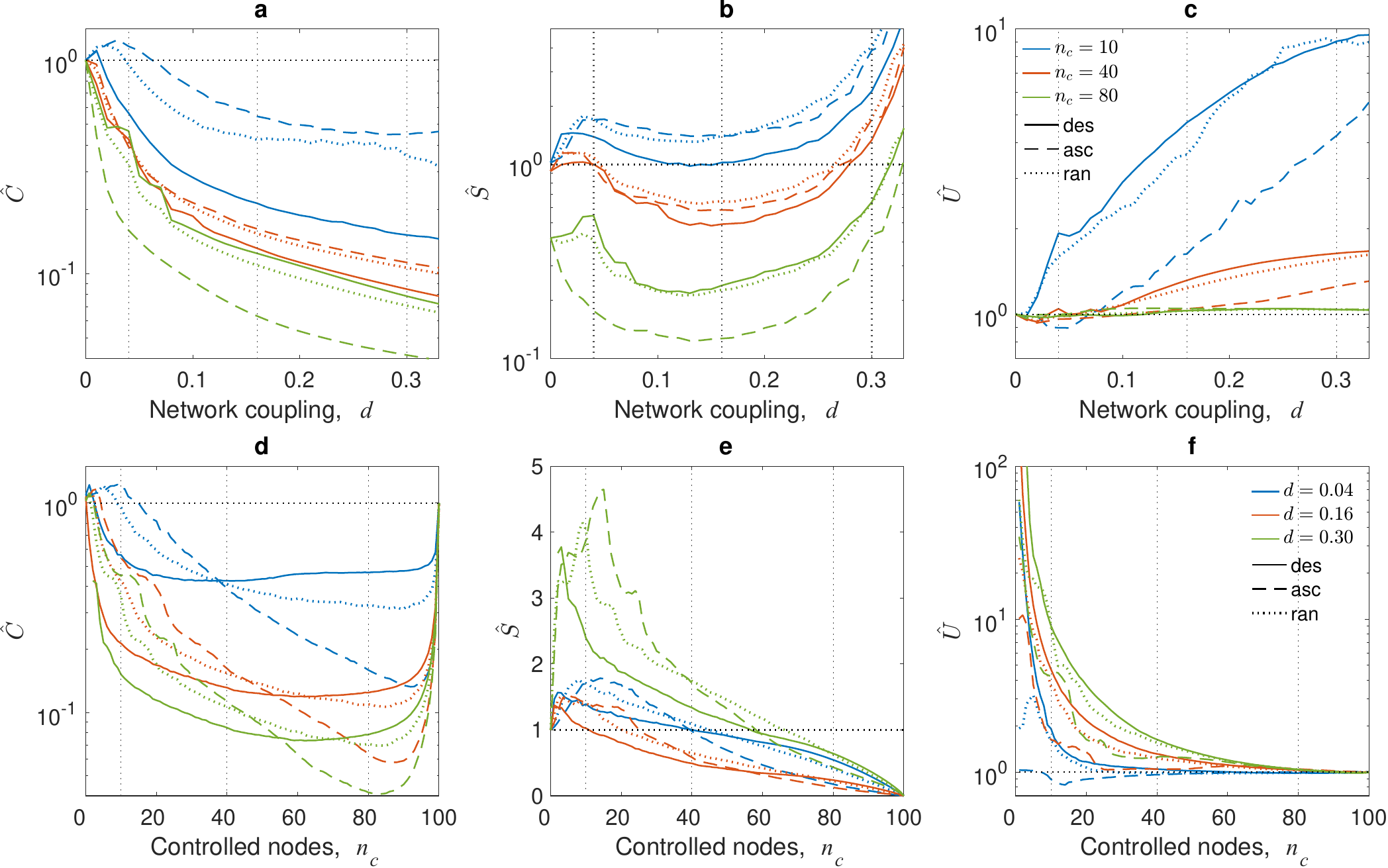}
	\caption{Normalized control error $\hat{C}$ (a, d), synchronization error $\hat{S}$ (b, e), and control energy $\hat{U}$ (c, f) extracted from the node-wise strategy panels in Fig.~\ref{fig:colormaps} along the one-dimensional cuts indicated there by white dashed lines, as functions of the coupling strength $d$ for fixed values of controlled nodes $n_c$ (top row panels) and as functions of $n_c$ for fixed $d$ (bottom row panels). Solid, dashed, and dotted curves correspond to node selection in descending  (des), ascending (asc), and random (ran) degree order, respectively. The random strategy, not displayed in Fig.~\ref{fig:colormaps}, is included here as a reference. Horizontal reference lines indicate the normalized level equal to one. Results are averaged over the same network realizations and initial conditions as in Fig.~\ref{fig:colormaps}.
\label{fig:cutscolormaps}
}
\end{figure*}

Figure~\ref{fig:colormaps} provides a global comparison among pairwise and 
node-wise control through the normalized synchronization error $\widehat{S}$,
control error $\widehat{C}$, and control energy $\widehat{U}$ as functions of 
the coupling strength $d$ and the number of controlled nodes $n_{\rm c}$. 
The black contours mark the performance thresholds specified in the caption, 
while the white dashed lines indicate the control parameter 
values selected for the one-dimensional sections shown in 
Fig.~\ref{fig:cutscolormaps}.

The two strategies exhibit distinct global performance patterns. For both of 
them, increasing the number of controlled nodes generally improves control and 
synchronization over an intermediate range of coupling strengths. However, 
node-wise control reaches low values of $\widehat{C}$ and $\widehat{S}$ with 
fewer controlled nodes and over a broader region of control parameter 
space. Pairwise control requires larger values of 
both $d$ and $n_{\rm c}$ to achieve comparable control accuracy and improves 
synchronization only within a comparatively restricted region. In both 
strategies, the synchronization error increases again at sufficiently strong 
coupling, indicating that the beneficial effect of coupling is not monotonic.

The clearest difference between the two strategies concerns how coupling 
modifies the control energy relative to the corresponding uncoupled reference. 
For pairwise control, $\widehat{U}$ becomes substantially larger than unity, 
particularly at weak coupling, although this increase decreases as $d$ grows. 
By contrast, the normalized energy of node-wise control remains much closer to 
$\widehat{U} \simeq 1$ throughout the explored parameter range. 
Hence, for node-wise control, the improvement in accuracy brought by network 
coupling comes at little additional energetic cost, except when only a few 
nodes are actuated at strong coupling. Because the two  quantities are 
normalized using architecture-specific reference configurations, these values 
characterize the coupling-induced change within each strategy rather than 
providing an absolute energy comparison between them.

The white regions in Fig.~\ref{fig:colormaps} identify control parameter 
combinations 
for which the controlled trajectories become unbounded and the performance
indicators cannot be evaluated. These regions occur predominantly for
pairwise control at weak coupling and represent the extension to large
networks of the low-coupling limitation already observed for the
controlled dyad in Fig.~\ref{fig:pair_synvscon}.

Figure~\ref{fig:cutscolormaps} examines the node-wise results in greater detail 
as a function of $n_{\rm c}$ and $d$, respectively. For 
clarity, the three node-selection strategies are compared simultaneously in
each panel: descending-degree, ascending-degree, and random selection
are represented by solid, dashed, and dotted lines, respectively.
The horizontal lines at unity indicate whether coupling improves or
degrades each performance measure relative to its corresponding
reference configuration.

The top row shows the dependence on the coupling strength for
$n_{\rm c} = \{ 10, 40, 80 \}$. In panel (a), the normalized control error
$\widehat{C}$ decreases as coupling becomes effective, showing that
network interactions enhance control. Moreover, increasing the number
of controlled nodes consistently produces lower control errors. Apart
from a small initial increase in the weak-coupling region, no
intermediate-coupling minimum followed by a systematic deterioration is
observed for $\widehat{C}$.

The behavior of the normalized synchronization error in panel (b) is
qualitatively different. For each value of $n_{\rm c}$, $\widehat{S}$ reaches
a minimum at an intermediate coupling strength and subsequently
increases. This increase should not necessarily be interpreted as a
deterioration of the absolute synchronization error $S$. Since
$\widehat{S}=S(n_{\rm c},d)/S(0,d)$, it reflects the fact that the
synchronization error of the uncontrolled network, $S(0,d)$, approaches
zero rapidly as coupling increases. The horizontal line
$\widehat{S}=1$ is therefore essential for the comparison: for
$n_{\rm c} = 10$, the curves remain predominantly above unity, indicating that
controlling only a small subset of nodes produces poorer intra-network
synchronization than in the uncontrolled network at the same coupling
strength. For $n_{\rm c} = 40$ and $80$, by contrast, broad intermediate
coupling intervals with $\widehat{S}<1$ appear, and synchronization
improves progressively as more nodes are controlled.

Panel (c) shows that, at fixed $n_{\rm c}$, the normalized control energy
generally increases with coupling. Nevertheless, for any given
coupling strength, the required energy decreases markedly as the
number of controlled nodes increases. When only a few nodes are
actuated, their controllers must compensate for the influence of many
uncontrolled neighbors. As more nodes approach the driver trajectory,
the local control errors entering the actuating signals become smaller,
thereby reducing the required control effort. Ascending-degree
selection consistently requires the least energy, while descending-degree and random selection exhibit remarkably similar energetic
costs.

The cuts at fixed coupling reveal that the most effective node-ranking
strategy depends on the available number of control inputs. When only
a small number of nodes can be actuated, descending-degree selection
provides the largest reduction of control and synchronization
errors, confirming the advantage of targeting hubs first. As $n_{\rm c}$
increases, however, the differences between the selection strategies
decrease and may eventually reverse: ascending-degree selection can
produce smaller errors and a lower control cost when a large fraction
of the network is directly actuated.\cite{Liu2021} Random selection generally
interpolates between these two limiting strategies. 

The bottom row also reveals how the number of controlled nodes interacts with 
the coupling strength. In panel (d), $\widehat{C}$ exhibits a minimum as a 
function of $n_{\rm c}$, most clearly for ascending-degree selection. As $d$ 
increases, this minimum shifts toward smaller values of $n_{\rm c}$, showing 
that stronger coupling reduces the number of direct control inputs required to
obtain the maximum network-mediated improvement in control. Beyond this 
optimum, $\widehat{C}$ increases again and approaches unity as $n_{\rm c} 
\rightarrow N$. This increase does not imply a deterioration of the absolute
control error $C$. Rather, when all nodes are controlled toward the same driver 
trajectory, their states become nearly identical, and the diffusive coupling 
terms vanish.  The coupled and uncoupled configurations then achieve
approximately the same control performance, $C(N,d)\simeq C(N,0)$, so that
$\widehat{C}(N,d)\simeq1$. At $n_{\rm c} = N$, the three selection strategies 
also become equivalent because they contain the same set of controlled nodes.

Panel (e) provides complementary evidence for an optimal intermediate
coupling strength. As $n_{\rm c}$ increases, the curves cross the reference
level $\widehat{S} = 1$ first for $d = 0.16$, then for $d = 0.04$, and finally
for $d = 0.30$. Thus, the intermediate value $d = 0.16$ requires the
smallest number of controlled nodes to improve synchronization relative
to the uncontrolled network, consistent with the minima observed in panel (b).

An analogous convergence is observed for the normalized control energy in panel 
(f), where all curves approach $\widehat{U} = 1$ as $n_{\rm c} \rightarrow N$. 
Once every node closely follows the driver, the differences between
neighboring states and the associated diffusive terms become negligible. The 
control energy is therefore essentially independent of the coupling strength,
$U(N,d)\simeq U(N,0)$, and hence $\widehat{U}(N,d) \simeq 1$. Thus, the 
convergence toward unity in both panels (d) and (f) reflects the disappearance
of the additional contribution of network coupling when the entire network is 
directly controlled.

\begin{figure*}
    \centering
	\includegraphics[width=\linewidth]{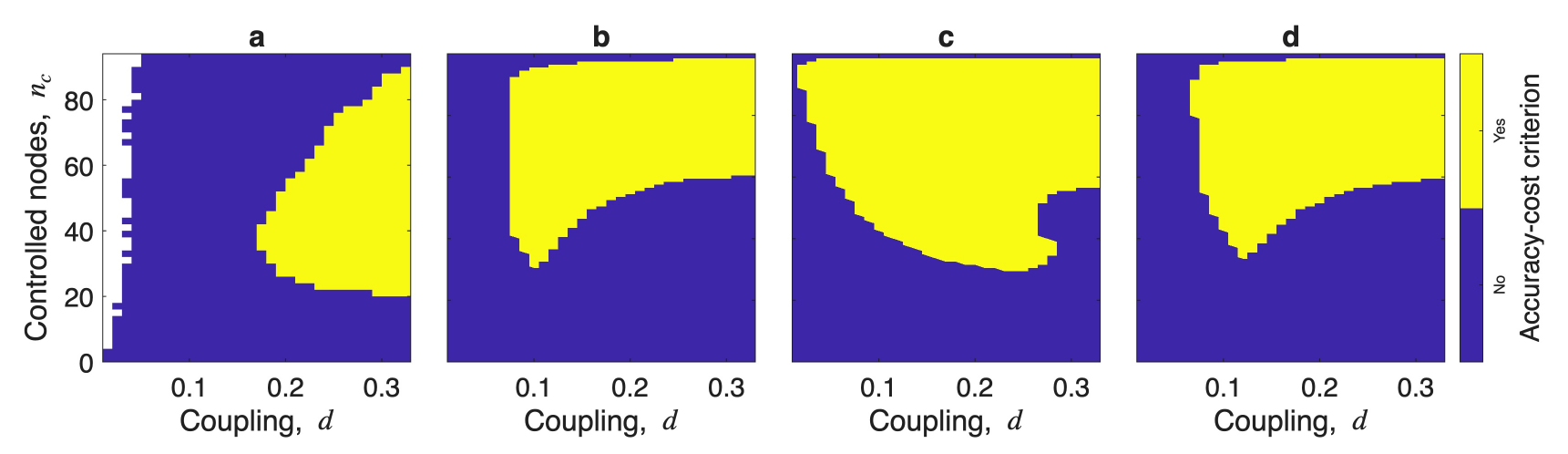} \\[-0.5cm]
    \caption{ Accuracy--cost operating regions obtained from the results shown 
in Fig.~\ref{fig:colormaps}. Panel (a) corresponds to pairwise control, while
panels (b)--(d) correspond to node-wise control with nodes selected in
descending degree order (b), ascending degree order (c), and random
order (d). For pairwise control, $n_{\rm c} = 2 n_{\rm p}$ denotes the 
equivalent number of controlled nodes. Yellow regions identify parameter
combinations satisfying simultaneously
$\widehat{C} < C_{\mathrm{th}} = 0.2$ and
$\widehat{U} < U_{\mathrm{th}}^{(\alpha)}$, with
$U_{\mathrm{th}}^{(\mathrm{pair})}=100$ and
$U_{\mathrm{th}}^{(\mathrm{node})} = 1.2$. Blue regions correspond to
bounded configurations that fail at least one condition, whereas white
regions identify configurations for which the controlled trajectories
become unbounded. 
    \label{fig:efficiency}}
\end{figure*}

To assess the trade-off between control accuracy and control effort within each 
control architecture, we identify the parameter combinations that 
simultaneously satisfy
\begin{equation}
   \widehat{C}(n_{\rm c},d) < C_{\mathrm{th}}, \qquad
	\widehat{U}(n_{\rm c},d) < U_{\mathrm{th}}^{(\alpha)},
\label{eq:accuracy-cost}
\end{equation}
where $\alpha \in \{\mathrm{pair},\mathrm{node}\}$. We use a common
control threshold because $\widehat{C}$ has the same interpretation
for both strategies. By contrast, the energy thresholds are
architecture-dependent because pairwise and node-wise control involve
different numbers and types of control channels.

Figure~\ref{fig:efficiency} shows the resulting accuracy--cost
operating regions for $C_{\mathrm{th}}=0.2$,
$U_{\mathrm{th}}^{(\mathrm{pair})}=100$, and
$U_{\mathrm{th}}^{(\mathrm{node})}=1.2$. Yellow indicates parameter
combinations satisfying both conditions, blue corresponds to bounded
configurations that fail at least one of them, and white identifies
configurations for which the controlled trajectories become unbounded.
The pairwise and node-wise maps should therefore be interpreted within
their respective energy scales.

For pairwise control [panel (a)], no accuracy--cost operating region
is found at weak coupling, where the control energy is large and some
trajectories become unbounded. A finite region emerges only above a
minimum coupling strength. Within this region, the lower boundary is
mainly determined by the control requirement, since too few controlled
pairs are insufficient to reduce $\widehat{C}$ below the prescribed
threshold. The upper boundary is instead imposed by the energetic
condition, so pairwise control satisfies both requirements only within
a restricted interval of $n_{\rm c}$ that broadens as the coupling becomes
stronger.

The three node-wise panels can be compared directly because they share
the same control architecture and the same accuracy and energy
thresholds. Among them, ascending-degree selection [panel (c)] produces
the largest accuracy--cost operating region over the sampled
$(d,n_{\rm c})$ plane. Provided that a sufficiently large number of nodes is
controlled, this region extends down to the smallest nonzero coupling
strength considered. Thus, no appreciable lower coupling threshold is
resolved for ascending-degree selection within the investigated
parameter range.

Descending-degree and random selection [panels (b) and (d)] yield
remarkably similar operating regions. In both cases, the simultaneous
accuracy and energy conditions are satisfied only above a finite
coupling strength, even when a large fraction of the network is
controlled. Their lower boundaries initially decrease with $d$, reach
a minimum at an intermediate coupling strength, and subsequently
increase again. Hence, stronger coupling does not indefinitely reduce
the number of actuators required to satisfy both criteria.

The larger region obtained for ascending-degree selection is consistent
with the energetic behavior observed in Fig.~\ref{fig:cutscolormaps}: 
controlling low-degree nodes first generally requires less control energy, 
allowing the energy condition to remain satisfied over a broader range of 
coupling strengths. This advantage becomes particularly relevant when many
nodes can be controlled. By contrast, descending-degree selection
remains preferable when the primary objective is to obtain accurate
control with a limited number of actuators, despite occupying a
smaller accuracy--cost operating region. Therefore, no single
node-selection strategy is optimal throughout the entire parameter
plane: the preferred ordering depends on the relative constraints on
actuator availability and control cost.

\subsection{Network-mediated propagation of control}\label{sec:network-mediation}

\begin{figure*}
	\includegraphics[width=\linewidth]{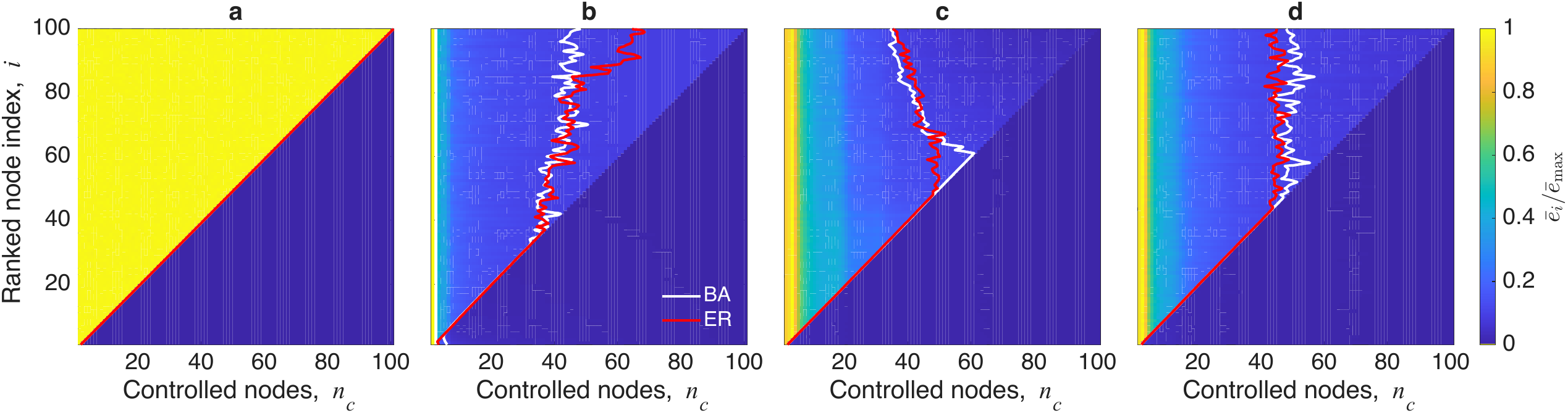} \\[-0.3cm]
    \caption{Node-level control performance under node-wise control.
Each panel shows the time-averaged control error
$\bar e_i(n_{\rm c},d)$ as a function of the number of actuated nodes
$n_{\rm c}$ (horizontal axis) and the node index $i$ in the corresponding
selection order (vertical axis). The background color maps show the
BA-network results normalized as
$\bar e_i/\bar e_{\max}$, where
$\bar e_{\max}=\max_{n_{\rm c},i}\bar e_i(n_{\rm c},d)$ is evaluated separately
for each panel. The diagonal $i=n_{\rm c}$ separates actuated
($i\leq n_{\rm c}$) and non-actuated ($i>n_{\rm c}$) nodes.
(a) Uncoupled reference, $d=0$; (b--d) coupled networks with
$d=0.23$ and nodes selected in descending-degree, ascending-degree,
and random order, respectively. White and red contours indicate
$\bar e_i/\bar e_{\max}=0.1$ for BA and ER networks, respectively,
with the normalization performed independently for each topology.
Deviations of these contours above the diagonal quantify the
network-mediated propagation of control to non-actuated nodes.
Networks have $N=100$ and $\langle k\rangle=4$. Results are averaged
over 100 independent network realizations and 5 initial conditions
per realization. Control parameters are $K_g=20$, $\lambda=1$, and
$\chi=0.7$.\label{fig:neteffect}}
\end{figure*}

The previous results indicate that network coupling contributes to
the spreading of the control action beyond the directly actuated
nodes. We now isolate this effect by analyzing the node-level
control error across the network. To assess the influence of degree
heterogeneity, we complement the BA networks considered above with
Erdős--Rényi (ER) random networks~\cite{Erdos1959}, constructed
with the same size, $N=100$, and mean degree, $\langle k\rangle=4$.

To make the network-mediated effect explicit,
Fig.~\ref{fig:neteffect} compares the node-level control errors in
the uncoupled reference configuration, $d=0$
[panel (a)], with those obtained for a coupled
network at $d=0.23$ [panels (b)--(d)], corresponding
to descending-degree, ascending-degree, and random node selection.
Each panel shows the normalized time-averaged control error of the
individual nodes as a function of the number of actuated nodes.
Nodes are arranged vertically according to their selection order,
so that the diagonal $i=n_{\rm c}$ separates directly actuated nodes from
those without actuators. The background color maps show the BA
results, normalized in each panel by
$\bar e_{\max}=\max_{n_{\rm c},i}\bar e_i(n_{\rm c},d)$. The white and red
contours mark the level $\bar e_i/\bar e_{\max}=0.1$ for BA and ER
networks, respectively, using the corresponding maximum for each
topology.

This comparison reveals a qualitatively different organization of
the node-level errors. In the uncoupled case, only the nodes that
directly receive a control input are driven toward the target
dynamics, producing a sharp separation along the diagonal. When
coupling is introduced, the low-error region extends into the
non-actuated part of the network, showing how nodes without
actuators approach the target through their connections to actuated
nodes. These node-level diagrams thus provide a direct visualization of 
network-mediated control: any low-error region above the diagonal corresponds 
to nodes that are controlled without being actuated.

More detailed information is provided by the shape of the white
(BA) and red (ER) contours. Under descending-degree selection, the
contours bend toward larger values of $n_{\rm c}$ as the node index
increases. Since the nodes selected later have progressively lower
degree, this indicates that peripheral nodes require a larger
actuated set before they can be effectively reached through the
network. The opposite trend appears under ascending-degree
selection: the contours bend toward smaller $n_{\rm c}$ because the high-degree nodes left without actuators retain multiple
pathways to the actuated part of the network and can therefore be
driven indirectly. For random selection, the contours remain approximately
vertical, consistently with the absence of a systematic relation
between node index and degree. These trends occur in both
topologies, although the broader degree distribution of BA networks
makes the dependence on node ranking more pronounced.

\begin{figure}
    \centering
    \includegraphics[width=0.4\textwidth]{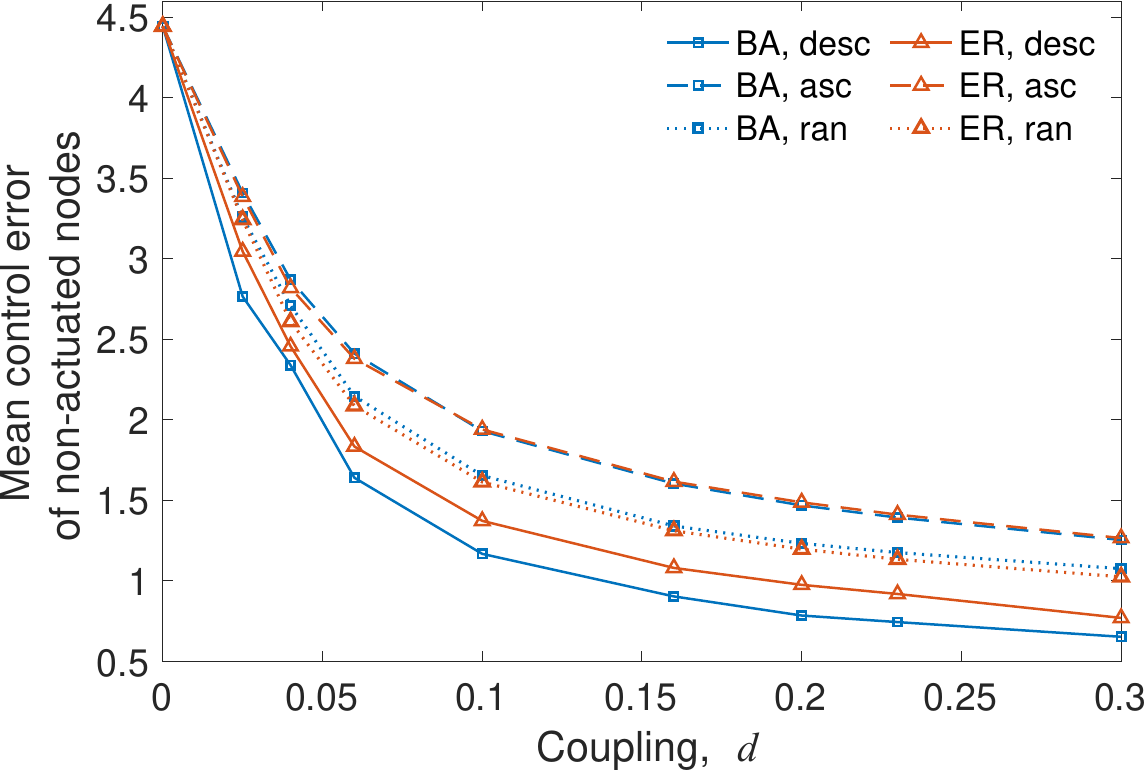}
    \caption{Mean control error of non-actuated nodes as a function of
the coupling strength $d$, for BA (blue) and ER (red) networks.
For each value of $d$, the average is taken over the non-actuated
nodes across all values of $n_{\rm c}$, corresponding to the region above
the diagonal $i=n_{\rm c}$ in the node-level maps of Fig.~\ref{fig:neteffect}.
Solid, dashed, and dotted lines correspond to node selection in
descending-degree, ascending-degree, and random order, respectively.
The decrease in the error with increasing coupling quantifies the
ability of the network to propagate the control action beyond the
directly actuated nodes. Control parameters and averaging conditions are as in Fig.~\ref{fig:neteffect}.\label{fig:averr_noctrl}}
\end{figure}

Since the normalization is performed independently for each panel
and topology, the contour geometry characterizes how propagation
depends on node rank rather than providing an absolute comparison
of control errors. Figure~\ref{fig:averr_noctrl} complements this
geometrical information by showing the mean, unnormalized control
error of non-actuated nodes as a function of coupling strength. For
each value of $d$, the average is taken over the non-actuated nodes
across the full range of $n_{\rm c}$ considered in
Fig.~\ref{fig:neteffect}.

For all selection strategies and network topologies, the mean error
decreases as coupling increases, confirming that the network itself
mediates the propagation of the control action. More importantly,
the separation between the curves shows that the effectiveness of
this propagation is governed by the connectivity of the actuated
nodes. Targeting hubs enhances the transmission of control,
particularly in the heterogeneous BA networks.

Together, these results show that the control signal does not act
only locally on actuated nodes. Instead, coupling allows the network
to transmit the control action to non-actuated nodes, and this
transmission is most effective when control inputs are placed on
highly connected nodes.

\section{Conclusions}

We generalize an observability-guided flatness control framework from isolated chaotic oscillators and dyads to heterogeneous networks beyond the dyadic setting. The proposed method determines measurement outputs and actuation channels directly from the nonlinear nodal dynamics. The resulting control laws are 
explicit, require only local measurements and their time derivatives, and do 
not rely on knowledge of the full network state. We consider two architectures: node-wise control, in which each selected oscillator is measured and actuated independently, and pairwise control, in which measurements from one node of an observable dyad are used to construct inputs applied to both nodes. 

The results reveal a trade-off between control performance and measurement-location requirements. Node-wise control generally achieves accurate control and synchronization over a broader parameter range, with its normalized control energy remaining closer to the corresponding uncoupled reference. Pairwise control requires only one measurement location per controlled dyad, but its control energy can increase significantly relative to its own reference, and the controlled trajectories may become unbounded at weak coupling.

Network coupling can propagate the effects of control to non-actuated nodes, creating a cooperative interaction between the collective dynamics and the external intervention. This effect depends on the coupling strength, the fraction and placement of actuated nodes, and the network heterogeneity. Targeting high-degree nodes enhances propagation when only a small number of nodes are actuated; as the controlled fraction increases, the advantages of degree-based placement diminish and can reverse. 

Overall, network interactions can reduce sensing needs and extend influence beyond actuated nodes. These results are demonstrated for networks of $N=100$ diffusively coupled R\"ossler oscillators and establish the applicability of the proposed architecture beyond isolated systems and single dyads. Assessing its performance for other nodal dynamics, network sizes, coupling functions, and topologies, as well as its robustness to noise, parameter mismatch, and delays, as well 
as centralized implementations of the pairwise scheme that account for
the cross-input terms neglected here, constitutes a natural direction for 
future work.

\section*{ACKNOWLEDGMENTS} I.S.N., I.L., and C.L. acknowledge financial support from the Spanish Ministerio de Ciencia, Innovación y Universidades under
Project  PID2023-147827NB-I00.

\bibliographystyle{aipnum4-2}
\bibliography{netcontrol}
\end{document}